\documentclass[prl,aps,twocolumn,amsmath,amssymb,showpacs,notitlepage]{revtex4-1}
\pdfoutput=1
\usepackage[dvipdfmx]{graphicx}
\usepackage{dcolumn}
\usepackage{bm}
\usepackage{color}
\usepackage{float}
\usepackage{wrapfig}
\usepackage{comment}
\usepackage{soul}
\begin{document}
\title{Microscopic Calculation of Spin Torques in Textured Antiferromagnets}
\date{\today}
\author{Jotaro J.~Nakane}
\affiliation{Department of Physics, Nagoya University, Nagoya 464-8602, Japan} 
\author{Hiroshi Kohno}
\affiliation{Department of Physics, Nagoya University, Nagoya 464-8602, Japan}
\begin{abstract}

 A microscopic calculation is presented for the spin-transfer torques (STT) and damping torques 
in metallic antiferromagnets (AF). 
 It is found that the sign of the STT is opposite to that in ferromagnets 
because of the AF transport character, 
and the current-to-STT conversion factor is enhanced near the AF gap edge.  
 The dissipative torque parameter $\beta_n$ and the damping parameter $\alpha_n$ 
for the N\'eel vector arise from spin relaxation of electrons.  
 Physical consequences are demonstrated for the AF domain wall motion using collective coordinates, 
and some similarities to the ferromagnetic case are pointed out such as intrinsic pinning and 
the specialty of $\alpha_n = \beta_n$. 
 A recent experiment on a ferrimagnetic GdFeCo 
near its angular-momentum compensation temperature is discussed.

\end{abstract}
\maketitle

 Manipulation of spin textures using electric current forms an intriguing subfield of spintronics.
 The effect of currents on ferromagnetic (FM) textures is 
well-understood through the spin angular momentum transfer between the conduction electrons 
and magnetization \cite{Ralph2008,Bazaliy,Brataas2012}. 
 However, a similar picture is not feasible in antiferromagnets (AFs) \cite{MacDonald2011,Duine2011,Jungwirth2016} 
since the magnetic order parameter and conduction electrons  
do not carry macroscopic spin angular momenta  \cite{Xu2008,Swaving2011,Hals2011,Tveten2013,Yamane2016,Park2020,Fujimoto2021}. 
 This makes microscopic studies indispensable for understanding spin torques in AFs.

 In FM, electrons moving in a spin texture with exchange coupling exhibit a spin polarization, 
\begin{align}
 \langle \hat {\bm \sigma} \rangle  \  \propto \   {\bm n} \times ({\bm v} \cdot\nabla) {\bm n} , 
\label{eq:sigma_STT} 
\end{align}
where ${\bm n}$ is the magnetic order parameter (magnetization vector for FM) 
and ${\bm v}$ is a velocity that characterizes the electron flow (spin current for FM). 
 The spin polarization arises as a reactive response \cite{Kohno_book} 
and exerts a reaction torque, known as the spin-transfer torque (STT), on the FM spins. 
 In AF, Xu {\it et al.} \cite{Xu2008} and Swaving and Duine \cite{Swaving2011} 
numerically obtained the same form of spin polarization as Eq.~(\ref{eq:sigma_STT}) with ${\bm n}$ 
now representing the N\'eel vector. 
 Analogous to FM, this spin polarization emerges through a reactive process, 
and gives rise to a torque that conserves total angular momentum, which may thus be called the STT. 
 However, in contrast to FM, the coefficient cannot be determined by a macroscopic argument 
based on the conservation law. 
 Moreover, there is in general another type of torque, called the $\beta$ torque, 
that arises as a dissipative response due to spin relaxation \cite{Zhang2004,Parkin2008}, 
the analytic expression of which is yet to be determined for AFs.

 In this Letter, we present a microscopic calculation of the STT, the $\beta$ torque, 
and the damping torques in AF metals. 
 A careful treatment is given to the effects of spin relaxation, 
which we model by magnetic impurities. 
 We find a STT proportional to the electric current but with a coefficient different from that in FM. 
 The $\beta$ torque is proportional to the spin-relaxation rate. 
 Interestingly, both torques in AFs drive the texture in the opposite direction 
compared to those in FMs.  
 Using collective coordinates, 
  it is shown that only the $\beta$ torque drives AF domain walls (DWs) \cite{Hals2011,Tveten2013}, 
because the effect of STT is nullified by an effect similar to the intrinsic pinning in FM. 
  Finally, a recent experiment on the current-assisted DW motion in ferrimagnets 
at the angular-momentum compensation temperature \cite{Okuno2019} is discussed.

 We consider the s-d model consisting of localized spins ($H_S$), 
conduction electrons ($H_{\rm el}$), 
and the s-d exchange interaction ($H_{\rm sd}$) between them, 
\begin{align}
    H &= H_S + H_{\rm el} + H_{\rm sd}  .
\end{align}
 The space dimensionality, $d$, can be arbitrary in the general formulation, 
but explicit calculations will be done for a two-dimensional square lattice, $d=2$.

 We first sketch the derivation of the equations that describe long-wavelength, 
low-frequency dynamics of AF spins coupled to conduction electrons. 
 We start with the lattice model,  
\begin{align}
  & H_S  = J\sum_{\langle i,j\rangle} {\bm S}_{i} \cdot {\bm S}_{j} - K\sum_{i} ({S}_i^z)^2  , 
\\
  & H_{\rm sd}  = - J_{\rm sd}\sum_i {\bm S}_{i} \cdot  c_i^\dagger {\bm\sigma} \, c_i  , 
\label{eq:Hsd}
\end{align}
where ${\bm S}_i$ is a localized, classical spin at site $i$, 
$J>0$ is the AF exchange coupling constant between nearest-neighbor (n.n.) sites 
$\langle i,j \rangle$, 
and $K>0$ is the easy-axis magnetic anisotropy constant. 
 In $H_{\rm sd}$, $c_i^\dagger = (c_{i\uparrow}^\dagger,c_{i\downarrow}^\dagger)$ 
is the electron creation operator at site $i$, 
${\bm \sigma}$ is a vector of Pauli matrices,  
and $J_{\rm sd}$ is the s-d exchange coupling constant.

 We consider a two-sublattice unit cell $m$ with localized spins, 
${\bm S}_{A,m}$ and ${\bm S}_{B,m}$, on each sublattice, 
and define the N\'eel component ${\bm n}_m$ and the uniform component ${\bm l}_m$ 
by \cite{Mikeska1991}
\begin{align}
    {\bm n}_m &= \frac{{\bm S}_{A,m}-{\bm S}_{B,m}}{2S}  , 
\ \ \ \ \   
    {\bm l}_m  = \frac{{\bm S}_{A,m}+{\bm S}_{B,m}}{2S}  , 
\label{eq:neel_uniform}
\end{align}
where $S = |{\bm S}_i|$ is the (constant) magnitude of the localized spins.  
 We assume the spatial variations are slow for ${\bm n}_m$ and ${\bm l}_m$, and 
adopt a continuum description, 
${\bm n}_m \to {\bm n} ({\bm r})$ and ${\bm l}_m \to {\bm l} ({\bm r})$. 
 We also exploit the exchange approximation, $|{\bm l}| \ll 1$, 
and neglect higher order terms in ${\bm l}$ \cite{Lifshitz_book}.  
 It then follows from 
 $| \, {\bm l} \pm {\bm n} | = 1$ that $ | {\bm n}| = 1$ and ${\bm n} \cdot {\bm l}= 0$.

 Since the definition in Eq.~(\ref{eq:neel_uniform}) violates sublattice-interchange 
symmetry \cite{Tveten2016}, 
it is convenient to work with the physical magnetization \cite{Nakane_AF_H},  
\begin{align}
  \tilde {\bm l} &\equiv  {\bm l} + \frac{a}{2}(\partial_x{\bm n}) , 
\label{eq:ell_physical} 
\end{align}
where $a$ is the lattice constant, 
and the $x$ axis is chosen along the bond connecting two sites in the unit cell. 
 This preserves the constraints, 
$\tilde {\bm l} \cdot {\bm n} = 0$ and $| {\bm n}| = 1$, 
within the exchange approximation, and simplifies the formalism. 
 In terms of $\tilde {\bm l}$ and ${\bm n}$, the Lagrangian density is given by \cite{Nakane_AF_H} 
\begin{align}
 \mathcal{L}_S  &= 
    s_n  \left\{ \, \tilde {\bm l} \cdot ({\bm n} \times \dot {\bm n})  
    - {\cal H}_{S} - {\cal H}_{\rm sd} \right\} , 
\label{eq:L_S}
\\
& {\cal H}_{S}
 = \frac{zJ S}{\hbar}  \bigg\{ \, \tilde {\bm l}^2  +  \frac{a^2}{4d} \sum_{i=1}^d (\partial_i {\bm n})^2
                         - \frac{K}{zJ} n_z^2  \bigg\} , 
\label{eq:H_S}
\\
& {\cal H}_{\rm sd}  
 = - \frac{M}{s_n} \, ( \, \tilde {\bm l} \cdot \hat {\bm \sigma}_l + {\bm n} \cdot  \hat {\bm \sigma}_n ) , 
\label{eq:H_sd}
\end{align}
where $s_n = 2\hbar S / (2a^d)$ is the density of staggered angular momentum, 
$z$ is the number of n.n.~sites of a given site, 
$\hat {\bm \sigma}_l$ and $\hat {\bm \sigma}_n$ 
are the uniform and staggered components of the electron spin density, and $M = J_{\rm sd}S$.  
 The equations of motion are obtained as  
\begin{align}
    \left\{
    \begin{array}{cl}
         \dot{\bm n}
&= \displaystyle
   {\bm H}_l  \times {\bm n}  + {\bm t}_n  , 
\\[8pt]
        \dot {\tilde {\bm l}}
&= \displaystyle
          {\bm H}_n  \times {\bm n}  + {\bm H}_l  \times \tilde {\bm l} + {\bm t}_{\, l}  , 
    \end{array}
    \right.  
\label{eq:afm_eom_sym}
\end{align}
where 
$ {\bm H}_n  = \partial \mathcal H_S / \partial {\bm n}$ 
and 
${\bm H}_l  = \partial \mathcal H_S/\partial \tilde {\bm l}$ are the effective fields  
coming from the spin part (${\cal H}_{S}$), and 
\begin{align}
 {\bm t}_n  
&= \frac{M}{s_n}   {\bm n} \times \langle \hat {\bm \sigma}_l \rangle , 
\\ 
 {\bm t}_{\, l} 
&= \frac{M}{s_n}  \left\{ \,  {\bm n} \times \langle \hat {\bm \sigma}_n  \rangle 
                                  + \tilde {\bm l} \times \langle \hat {\bm \sigma}_l \rangle  \right\}   , 
\end{align}
are the spin torques from electrons (${\cal H}_{\rm sd}$).  
 We calculate 
$\langle \hat {\bm \sigma}_l \rangle$ and $\langle \hat {\bm \sigma}_n \rangle$ 
in response to an applied electric field ${\bm E}$ or to the time-dependent ${\bm n}$ and $\tilde {\bm l}$
using the Kubo formula \cite{Kohno2006,Kohno2007}.

 To be explicit, we consider tight-binding electrons on a two-dimensional square lattice 
described by 
\begin{equation}
	H_{\rm el} =  
	- t \sum_{\langle i,j\rangle} (c_{i}^\dagger c_j + {\rm h.c.}) 
    + V_{\rm imp} , 
\end{equation}
where the first term expresses n.n.~hopping, and 
\begin{align}
 V_{\rm imp} &= 
	u_{\rm i} \sum_l c_l^\dagger c_l 
+   u_{\rm s} \sum_{l'} {\bm S}_{l'}^{\rm imp} \cdot c_{l'}^\dagger {\bm\sigma} \, c_{l'} , 
\end{align}
defines coupling to nonmagnetic and magnetic impurities. 
 Combined with $H_{\rm sd}$, the hopping term gives upper and lower (spin-degenerate) bands, 
$\pm E_{\bm k} = \pm \sqrt{\varepsilon_{\bm k}^2 + M^2}$, in a uniform AF state, 
where $\varepsilon_{\bm k} = -2t (\cos k_x + \cos k_y)$. 
 We take a directional average of ${\bm S}_j^{\rm imp}$ 
with second moment $\overline{S_z^2}$ ($\overline{S_\perp^2}$) 
for the component parallel (perpendicular) to ${\bm n}$. 
 In the Born approximation, they appear through 
$\gamma_{\rm n} = \pi n_{\rm i}u_{\rm i}^2 \nu$,  
$\gamma_\perp = \pi n_{\rm s} u_{\rm s}^2 \, \overline{S_\perp^2} \, \nu$, and 
$\gamma_z = \pi n_{\rm s} u_{\rm s}^2 \, \overline{S_z^2} \, \nu$,  
where $n_{\rm i}$ and $n_{\rm s}$ are the respective impurity concentrations, 
and $\nu = \frac{1}{N} \sum_{\bm k} \delta(|\mu|-E_{\bm k})$  
is the density of states per spin ($N$ is the total number of sites) 
with the chemical potential $\mu$ measured from the AF gap center.

 Vertex correction is necessary for a proper account of spin conservation, or its weak violation. 
 Here, it is evaluated in the ladder approximation, 
\begin{align}
    \Pi_{\sigma\bar{\sigma}}
&= \frac{2}{\pi\nu\tau^2} \frac{\mu^2}{\mu^2-M^2} \, 
    \frac{1}{Dq^2 - i\omega + \tau_\varphi^{-1} + \tau_{\rm s}^{-1} } . 
\label{eq:Pi_0}
\end{align}
 This describes diffusion, dephasing, and relaxation of transverse spin density, 
generalizing the result of Ref.~\cite{Nakane2020} to include magnetic impurities.  
 Here, $\tau^{-1} = 2 \, [ \gamma_+ + (M/\mu)^2 \gamma_- ]$, with 
$\gamma_\pm = \gamma_{\rm n} + \gamma_z \pm 2\gamma_\perp$,  
is the electron scattering rate,  and 
\begin{align}
  \frac{1}{\tau_\varphi}  
&= \frac{4M^2}{\mu^2 } \left[ 
     \frac{\mu^2+M^2}{\mu^2-M^2} \gamma_{\rm n} 
    + 3 \gamma_\perp + \frac{2(2\mu^2+M^2)}{\mu^2-M^2} \gamma_z  \right] , 
\label{eq:tau_phi}
\\
  \frac{1}{\tau_{\rm s}}  &= 4 \, (\gamma_\perp + \gamma_z  ) , 
\end{align}
are, respectively, the spin-dephasing rate \cite{Manchon2017,Nakane2020} and 
the (transverse) spin-relaxation rate. 
 In Eq.~(\ref{eq:Pi_0}), 
$q^{-1}$ ($\omega^{-1}$) is the typical length (time) scale of the AF spin texture (dynamics), 
and $D$ is the diffusion constant. 
 We assume $q \ell_\varphi \ll 1$ and $\omega \tau_\varphi \ll 1$, 
where $\ell_\varphi = \sqrt{D \tau_\varphi}$ is the spin-dephasing length, 
and let $q, \omega \to 0$ in the results. 
 The constant terms in the denominator, $\tau_\varphi^{-1} + \tau_{\rm s}^{-1}$,  
reflect spin nonconservation in the electron system. 
 The spin dephasing ($\tau_\varphi^{-1}$), characteristic of AF and absent in FM, 
is dominated by nonmagnetic impurities and vanishes at $M = 0$ \cite{Manchon2017}, 
whereas $\tau_{\rm s}^{-1}$ comes solely from magnetic impurities 
and is essentially the same as that in FM \cite{Kohno2006}. 
 It is convenient to decompose the former as 
$\tau_\varphi^{-1} =\tau_{\varphi 0}^{-1} + \tau_{\varphi 1}^{-1}$, 
where $\tau_{\varphi 0}^{-1}$ ($\propto \gamma_{\rm n}$) is the contribution from nonmagneic impurities     
and $\tau_{\varphi 1}^{-1}$ is from magneic impurities. 
 The \lq\lq dissipated'' spin angular momentum via $\tau_{\varphi 0}^{-1}$  
is actually transferred to the AF spin system.

 We calculate electron spin density induced by an external electric field ${\bm E}$ 
in the presence of spin texture (for current-induced torques), 
or induced by time-dependent spins, $\dot {\bm n}$ and $\dot {\tilde {\bm l}}$ (for Gilbert damping). 
 We assume weak spin relaxation, $\gamma_z,  \gamma_\perp \ll \gamma_{\rm n}$, 
and retain terms of lowest nontrivial order. 
 The calculations are straightforward along the lines of 
Refs.~\cite{Kohno2006,Kohno2007,Nakane2020}; see \cite{Suppl} for details.

{\it Results.}--- We obtained 
\begin{align}
\langle \hat {\bm \sigma}_n \rangle 
&= - (s_n/M) \{ \beta_n \, ({\bm v}_n \!\cdot\! \nabla) \, {\bm n} 
      + \alpha_n \dot{\bm n}  \} , 
\label{eq:sigma_n} 
\\
 \langle \hat {\bm \sigma}_l \rangle 
&= (s_n/M) \{ {\bm n} \times ({\bm v}_n \!\cdot\! \nabla) \, {\bm n} 
     - \alpha_l \, \dot {\tilde {\bm l}} \} , 
\label{eq:sigma_l} 
\end{align}
which are consistent with previous studies
 \cite{Xu2008,Swaving2011,Hals2011,Tveten2013,Yamane2016,Park2020}, 
and lead to the torques, 
\begin{align}
 {\bm t}_n &= - ({\bm v}_n \!\cdot\! \nabla) {\bm n} 
    - \alpha_l  \, {\bm n} \times  \dot {\tilde {\bm l}}  , 
\label{eq:t_n} 
\\
 {\bm t}_{\, l} &= - \beta_n \, {\bm n} \times ({\bm v}_n \!\cdot\! \nabla) {\bm n} 
    - \alpha_n \, {\bm n} \times  \dot {\bm n} . 
\label{eq:t_l} 
\end{align}
 We retained dominant contributions, which come from 
$\langle \hat {\bm \sigma}_l \rangle$ for ${\bm t}_n$, 
and $\langle \hat {\bm \sigma}_n \rangle$ for ${\bm t}_{\, l}$.

{\it  Current-induced torques.}---
 The first terms in ${\bm t}_{\, l}$ and ${\bm t}_n$ are current-induced torques, 
which are proportional to the charge current density, 
${\bm j} = \sigma_{xx}{\bm E} = 2e^2 D\nu{\bm E}$, via 
\begin{align}
 {\bm v}_n = -  \frac{\hbar}{2e s_n} {\cal P}_n {\bm j} . 
\end{align}
 The velocity ${\bm v}_n$ quantifies the STT, and we identify 
\begin{align}
 {\cal P}_n =  \frac{\mu M}{\mu^2-M^2}  , 
\label{eq:P_n} 
\end{align}
to be the conversion factor from a charge current to STT. 
 Note that $|{\cal P}_n |$ can be significantly larger than unity near the AF gap edge ($|\mu| \gtrsim |M|$). 
 This contrasts to the case of FM, in which the corresponding factor $|P|$ is less than unity. 
 The current-induced torque in ${\bm t}_{\, l}$ is characterized by a dimensionless parameter, 
\begin{align}
    \beta_n &=  \frac{2 (\gamma_\perp + \gamma_z )}{M}  
= \frac{\hbar}{2 M \tau_{\rm s}}  , 
\label{eq:beta_n}
\end{align}
which originates from magnetic impurities, i.e., from spin relaxation, 
and is therefore a dissipative torque.
 The spin dephasing due to nonmagnetic impurities ($\tau_{\varphi \, 0}^{-1}$) 
is microscopically a reactive process and does not contribute to $\beta_n$, 
whereas that from $\tau_{\varphi \, 1}^{-1}$, combined with the self-energy terms,  
results in a contribution proportional to $\tau_{\rm s}^{-1}$. 
 Along with the contribution originating from $\tau_{\rm s}^{-1}$ in Eq.~(\ref{eq:Pi_0}), 
it leads to Eq.~(\ref{eq:beta_n}). 
 The obtained two current-induced torques are related via $\beta_n{\bm n} \times$, 
which is reminiscent of the relation between the reactive and dissipative torques in FM; 
we call the former [$-({\bm v}_n \!\cdot\! \nabla){\bm n}$] the STT in AF, 
and the latter [$-\beta_n {\bm n} \times ({\bm v}_n \!\cdot\! \nabla){\bm n}$] the $\beta_n$ torque. 
 The expression of $\beta_n$ in terms of $\tau_{\rm s}$ 
and $M = J_{sd} S$ is also shared by FM \cite{Kohno2006,Zhang2004}.

 The above two current-induced torques change their signs across the AF gap [see Eq.~(\ref{eq:P_n})], 
reflecting the fact that electrons in the upper and lower AF bands have mutually opposite spin directions.  
 This feature of the STT was suggested in Ref.~\cite{Swaving2011}. 
 Interestingly, the driving direction is opposite to the naive expectation 
based on the two-FM picture of AF. 
 Namely, for $\mu < 0$, the spin polarization on the Fermi surface is positive 
(dominated by majority spin carriers) but the driving direction is opposite 
to the direction of electron flow. 
This is due to the intersublattice hopping in AF, namely, 
the electron spins exert torques on oppositely pointing neighboring spins, 
so the sign of the torques is reversed from that of FMs \cite{com_nn}.
 The same is true for $\mu > 0$.

{\it Gilbert damping.}---
 The second terms in Eqs.~\eqref{eq:t_n} and \eqref{eq:t_l} describe damping.    
 The damping parameters are calculated as 
\begin{align}
 \alpha_n 
  &=   \left\{    \gamma_\perp + \gamma_z  
             + \frac{M^2}{\mu^2} ( \gamma_\perp - \gamma_z )  \right\} \frac{2\hbar \nu }{s_n} , 
\\
  \alpha_l &= \frac{(\mu^2-M^2)(\mu^2+M^2)}{\mu^2}  \frac{\nu}{s_n} \tau . 
\label{eq:alpha_AF}
\end{align}
 While $\alpha_n$ arises from spin relaxation (magnetic impurities), $\alpha_l$ does not necessitate it. 
 Rather, $\alpha_l$ is proportional to $\tau$, like conductivity, hence can be very large 
in good metals. 
 These features were pointed out in Refs.~\cite{Liu2017,Simensen2020} 
based on the first-principles calculation.

 It is interesting to compare $\alpha_n$ with the Gilbert damping in FM, 
\begin{align}
 \alpha_{\rm F} &= \sum_\sigma 
  (\gamma_{z,\sigma}\nu_{\bar\sigma} +\gamma_{\perp,\sigma}\nu_{\sigma} ) \, \frac{\hbar}{ s_0 }, 
\label{eq:alpha_F}
\end{align}
obtained based on the same spin-relaxation model (magnetic impurities) \cite{Kohno2006}. 
 Here, 
$\gamma_{\alpha, \sigma} = \pi n_s u_s^2 \overline{S^2_\alpha} \, \nu_\sigma$
($\alpha = \perp, z$), $\nu_\sigma$ ($\sigma = \uparrow, \downarrow$) 
is the density of states of electrons with spin $\sigma$, 
and $s_0 = \hbar S/a^d$ is the angular-momentum density. 
 We see that in the limit of spin-degenerate bands ($\nu_\uparrow = \nu_\downarrow$) 
and isotropic magnetic impurities ($\gamma_\perp = \gamma_z$), 
 the above expressions of $\alpha_n$ (for AF) and $\alpha_{\rm F}$ (for FM) coincide.
 Therefore, in the current model of AF, the ratio $\beta_n /\alpha_n$ is of order unity, 
similar to FM \cite{Kohno_book}.

{\it  Equations of AF spin dynamics.}---
 With the obtained torques and ${\cal H}_{S}$ [Eq.~(\ref{eq:H_S})], 
the equations of motion are explicitly written as 
\begin{align}
   \dot{\bm n} 
&=  \tilde J \,  \tilde {\bm l}  \times {\bm n} - ({\bm v}_n \!\cdot\! \nabla ) \, {\bm n}  , 
\label{eq:eom_n1}
\\
   \dot {\tilde {\bm l} } 
&=   - ( c^2 \tilde J^{-1} \nabla^2{\bm n} + \tilde K  n^z \hat{z}  )  \times{\bm n}
\nonumber \\
&\quad
        + ( \alpha_n  \dot{\bm n} + \beta_n ({\bm v}_n \!\cdot\! \nabla ) \, {\bm n})  \times {\bm n}  
\nonumber \\
&\quad  + {\bm n} \, [\, \tilde {\bm l} \!\cdot\! ({\bm v}_n \!\cdot\! \nabla){\bm n}]  ,  
\label{eq:eom_l1}
\end{align}
with $c = (zJSa) / (\hbar\sqrt{d})$, $\tilde J =  2zJS /\hbar $, and $\tilde K = 2SK /\hbar $. 
  Damping terms in the first equation are dropped as they are higher order in $\tilde {\bm l}$. 
 Solving Eq.~\eqref{eq:eom_n1} for $\tilde {\bm l}$ as 
$ \tilde {\bm l}  =
    \tilde J^{-1}  {\bm n} \times [\dot{\bm n} + ({\bm v}_n \!\cdot\! \nabla ) \,  {\bm n} ]$, 
and substituting it in Eq.~\eqref{eq:eom_l1}, one can obtain a closed equation for ${\bm n}$, 
\begin{align}
 \ddot {\bm n} \times {\bm n} 
&= ( c^2 \, \nabla^2{\bm n}   + \tilde J \tilde K n_z \hat{z} )  \times {\bm n}
\nonumber\\
&\ \ \ 
    - \tilde J 
   ( \alpha_n  \dot{\bm n} + \beta_n ({\bm v}_n \!\cdot\! \nabla ) \, {\bm n})  \times {\bm n} 
\nonumber\\
&\ \ \   - [ ({\bm v}_n \!\cdot\! \nabla ) \, \dot {\bm n} ] \times {\bm n}  . 
\label{eq:eom_n2}
\end{align}
 This differs slightly from Ref.~\cite{Swaving2011} due to the difference  in $H_{\rm sd}$ 
(i.e., ${\bm l}$ vs.~$\tilde {\bm l}$), and leads to the magnon dispersion, 
\begin{align}
 \omega &= \sqrt{c^2 {\bm q}^2 + \tilde J \tilde K 
  + ({\bm v}_n \!\cdot\! {\bm q} - i \tilde J \alpha_n)^2 /4 
  + i \tilde J \beta_n {\bm v}_n \!\cdot\! {\bm q}}
\nonumber \\
& \ \ \   \pm ({\bm v}_n \!\cdot\! {\bm q} - i \tilde J \alpha_n )/2  , 
\end{align}
where damping enters only through $\alpha_n$ and $\beta_n$.

{\it  DW motion.}---
 Here we study the AF DW motion using collective coordinates. 
 Since $\mathcal{L}_S$ [Eq.~(\ref{eq:L_S})] is written with ${\bm n}$ and $\tilde {\bm l}$,
 we consider collective coordinates for both ${\bm n}$ and $\tilde {\bm l}$ \cite{Nakane_AF_H}.
 Assuming for ${\bm n} = (\sin\theta \cos\phi, \sin\theta \sin\phi, \cos\theta)$ a DW form,  
$\cos\theta(x,t) = \pm\tanh \left( \frac{x-X(t)}{\lambda} \right)$ and 
$\phi (x,t) = \phi_0 (t)$, 
where $\lambda =  a \sqrt{zJ/4Kd}$ is the DW width, we treat the DW position 
$X(t)$ and the angle $\phi_0 (t)$ as dynamical variables \cite{Tatara2008}. 
 As for $\tilde {\bm l}$, we expand it as \cite{Nakane_AF_H} 
\begin{align}
   \tilde  {\bm l}(x,t)
    &=   [ \, l_\theta(t) \, {\bm e}_{\theta} + l_\phi(t) \, {\bm e}_{\phi} \, ] \, \varphi_0(x) + \cdots ,
\end{align}
where ${\bm e}_\theta \equiv \partial_\theta {\bm n}$ and 
${\bm e}_\phi \equiv {\bm n} \times{\bm e}_\theta$ are orthonormal vectors normal to ${\bm n}$. 
 The function $\varphi_0 (x) =  \Bigl[ \cosh\frac{x-X}{\lambda} \Bigr]^{-1}$  
reflects the spatial profile of ${\bm n} \times \dot {\bm n}$, 
and naturally extracts $l_\theta$ and $l_\phi$ in the first term 
of $\mathcal{L}_S$ [Eq.~(\ref{eq:L_S})]. 
 The obtained Lagrangian, 
$L_{\rm DW} = 2s_n  (\pm \dot X l_\phi - \lambda \dot \phi_0 l_\theta ) - H_S$, 
shows that $l_\phi$ and $l_\theta$ are canonical conjugate to $X$ and $\phi_0$, 
respectively. 
 The equations of motion are given by 
\begin{align}
     \pm   \lambda \, \dot{l}_\phi  &= \beta_n v_n - \alpha_n \dot{X}  , 
\label{eq:eom_x}
\\%%%%%%%%%%%%%%%%%%%%%%%%%%%%%%%%%%%%
     \pm \dot{X}   &=  \pm v_n + v_J \, l_\phi  + \alpha_l \lambda  {\dot l}_\phi     , 
\label{eq:eom_lphi}
\\%%%%%%%%%%%%%%%%%%%%%%%%%%%%%%%%%%%%
    \dot l_\theta &= \alpha_n   \, \dot{\phi}_0  , 
\label{eq:eom_phi}
\\%%%%%%%%%%%%%%%%%%%%%%%%%%%%%%%%%%%%
    \lambda \, \dot{\phi}_0  &= -  v_J  l_\theta - \alpha_l \lambda  \dot l_\theta      , 
\label{eq:eom_ltheta}
\end{align}
where $v_J = 4 J S \lambda /\hbar$, and $\pm$ is the topological charge of the AF DW. 
 The first two equations describe the translational motion, 
and the remaining two describe the rotational motion of the DW plane (defined by the N\'eel vector). 
 Unlike in FM \cite{Tatara2004}, these two motions are decoupled in AF. 
 The term $\pm v_n$ in Eq.~(\ref{eq:eom_lphi}) describes the spin-transfer effect,  
and $\beta_n v_n$ in Eq.~(\ref{eq:eom_x}) describes the momentum-transfer effect 
(a force on the DW). 
 The terms with $\alpha_l$ are negligible in effect, 
but retained here for the sake of comparison with FM (see below).

 Let us overview the translational motion of AF DW under a stationary 
$v_n$ \cite{Hals2011}. 
 When $\alpha_n = \beta_n = 0$, $l_\phi$ is a constant of motion. 
 With an initial condition $l_\phi = 0$ (no canting), 
 the DW moves at a constant velocity $\dot{X} = v_n$ by the spin-transfer effect \cite{Swaving2011}.  
 If the DW is initially canted, $l_\phi = l_\phi^0$, the constant velocity is modified to 
$\dot{X} = v_n \pm v_J \, l_\phi^0$. 
 For finite $\alpha_n$, $l_\phi$ is no longer conserved, and approaches a terminal value, 
$ l_\phi  \to \mp ( 1 - \beta_n/\alpha_n ) (v_n/v_J)$. 
 Then, from Eq.~(\ref{eq:eom_lphi}), the DW velocity approaches
\begin{align}
  \dot{X}   \  \to  \    v_n  -  \left( 1 - \frac{\beta_n}{\alpha_n} \right)  v_n  
= \frac{\beta_n}{\alpha_n} \, v_n  , 
\end{align}
which is solely determined by the $\beta_n$ torque. 
 If $\beta_n = 0$, the spin-transfer effect  is completely nullified by the canting 
$l_\phi = v_n/v_J$, 
and the aforementioned steady movement eventually ceases \cite{Hals2011}. 
 This is quite similar to the intrinsic pinning in FM. 
 For finite $\beta_n$, canting $l_\phi$ is reduced, 
and the cancellation of the spin-transfer effect is incomplete. 
 Finally, the case $\beta_n = \alpha_n$ is special in that there is no canting, and 
the spin-transfer effect is undisturbed.

 It is instructive to make a more detailed comparison with FM. 
 In FM, the current-driven DW motion is described by 
\begin{align}
    \pm \lambda \, \dot \phi_0  &= \beta v_{\rm s} - \alpha \dot X , 
\label{eq:eom_x_F}
\\
    \pm \dot X   &=  \pm v_{\rm s} + v_K  \sin 2 \phi_0 + \alpha \lambda \dot \phi_0  , 
\label{eq:eom_phi_F}
\end{align}
where, now, $X$ and $\phi_0$ are coupled. 
 ($\phi_0$ here is defined by the uniform magnetization, 
and $\pm$ is the topological charge of the FM DW.) 
 A close similarity to Eqs.~(\ref{eq:eom_x}) and (\ref{eq:eom_lphi}) is evident, 
and here $\phi_0$ plays the role of $\l_\phi$. 
 The effect of current appears in $v_{\rm s} = -(\hbar/2es_0) P j$,  
where $P$ is the current polarization factor, 
and the velocity $v_K = K_\perp S \lambda /2 \hbar$ is defined 
with the hard-axis anisotropy constant $K_\perp$. 
 At low current,  $v_{\rm s} < v_K$, and with $\beta=0$, 
the DW plane tilts by $\phi_0 = (1/2) \sin^{-1} (v_{\rm s}/v_K)$ 
and the DW ceases to move, $\dot X = 0$. 
 This is the intrinsic pinning in FM  \cite{Tatara2004,Koyama2011}. 
 If $v_{\rm s}$ exceeds $v_K$, $v_K$ can not nullify the spin-transfer effect  $v_{\rm s}$ 
 and the DW is released from intrinsic pinning. 
 The corresponding term in Eq.~(\ref{eq:eom_lphi}) has the linearised form, $v_J \, l_\phi $,  
which is justified since $v_J$ of AF is much larger than $v_K$ of FM 
(by 2-3 orders of magnitude), and the intrinsic pinning is robust in AF. 
 It is interesting to note the contrasting origins of intrinsic pinning; 
in FM it is the explicit breaking of spin rotation symmetry, $K_\perp$, 
whereas in AF it is the AF order itself, i.e., spontaneous breaking. 
 Finally, the case $\alpha = \beta$ provides a special solution $\phi_0 = 0$ and $\dot{X} = v_s$, 
similar to the case $\alpha_n = \beta_n$ for AF.

 Recently, current-assisted field-driven DW motion was experimentally studied 
in a {\it ferrimagnetic} GdFeCo near its angular-momentum compensation temperature \cite{Okuno2019}. 
 They analyzed the data by the Landau-Lifshitz-Gilbert equation 
for the {\it uniform moment} ${\bm m}$ (parallel to ${\bm n}$), 
and obtained a very large, negative value of $\beta / \alpha  \simeq -100$. 
 They assumed $\sim \beta P {\bm j}$ for the $\beta$-torque coefficient (that acts on ${\bm m}$),  
with a small factor $P$ $(\simeq 0.1)$ included. 
 If, however, the main driving is the $\beta_n$ torque that acts 
on the {\it N\'eel vector} ${\bm n}$, as studied in the present work, 
we would conclude ${\cal P}_n \beta_n/\alpha_n \simeq -10$. 
 While $\beta_n/\alpha_n \simeq 1$ as in FM (for positive $J_{\rm sd}$ \cite{Hoshi2020,com2}), 
$|{\cal P}_n|$ can be significantly larger than unity 
near the AF gap edge. 
 Therefore, the large value of $|{\cal P}_n| \sim 10$ may lie within the scope of the present results.  
 The negative sign can be explained likewise from ${\cal P}_n$ with a negative $\mu$, 
which reflects intersublattice hopping in AF. 
 Such \lq\lq antiferromagnetic transport'' in GdFeCo is supported by a recent magnetoresistance 
measurement \cite{Park2021}.

 In conclusion, we have presented a microscopic model calculation of current-induced torques 
and damping torques in AF metals, 
paying attention to the effects of spin relaxation (and spin dephasing). 
 A formulation in terms of the N\'eel vector and physical magnetization 
is given to study the AF spin dynamics in metallic AFs with s-d exchange interaction.  
 The current-induced torques are found to be opposite in direction to those of FMs, 
reflecting the AF transport character, 
and the current-to-STT conversion factor can be significantly larger than that in FM. 
 These results seem to be relevant to the recent experiment on GdFeCo.

 We would like to thank T. Okuno, T. Ono and T. Taniyama for valuable discussions. 
 We also thank K. Nakazawa, T. Funato, Y. Imai, T. Yamaguchi, A. Yamakage, and Y. Yamazaki for helpful discussion. 
 This work was partly supported by JSPS KAKENHI Grant Numbers JP15H05702, JP17H02929 and JP19K03744, 
and the Center of Spintronics Research Network of Japan. 
 JJN is supported by a Program for Leading Graduate Schools ``Integrative Graduate Education and Research in Green Natural Sciences'' 
 and Grant-in-Aid for JSPS Research Fellow Grant Number 19J23587.


\begin{references}
\bibitem{com_nn} We have confirmed this by introducing the next n.n. hopping $t'$. 
 As $t'$ starts dominating, the sign of the STT 
changes from negative to positive in the $\mu < 0$ region, and vice versa for $\mu > 0$, 
in conformity with the change of transport character from AF like to FM like \cite{Nakane_next}.  
 Note that the torques in FM also change their signs in a tight-binding model 
as $\mu$ is increased from the band bottom to the top 
since the majority/minority spins are switched and the sign of $P$ is reversed.
\end{references}
\end{document}

\title{Supplemental Material  for \\ 
 \lq\lq Microscopic Calculation of Spin Torques in Textured Antiferromagnets''}
\date{\today}
\author{Jotaro J.~Nakane}
\affiliation{Department of Physics, Nagoya University, Nagoya 464-8602, Japan} 
\author{Hiroshi Kohno}
\affiliation{Department of Physics, Nagoya University, Nagoya 464-8602, Japan}
\begin{abstract}

\end{abstract}
\maketitle

\section{Model of electrons}

 The conduction electrons are considered in the tight-binding Hamiltonian
and s-d exchange interaction, 
\begin{align}
	H_{\rm el} &=  
	- t \sum_{\langle i,j\rangle} (c_{i}^\dagger c_j + {\rm h.c.}) 
    + V_{\rm imp} , 
\\
    H_{\rm sd}&=-J_{\rm sd}\sum_i {\bm S}_{i} \cdot  c_i^\dagger {\bm\sigma} \, c_i
\end{align}
where  $c_i^\dagger = (c_{i\uparrow}^\dagger,c_{i\downarrow}^\dagger)$ is the electron creation operator at site $i$. 
 The first term in $H_{\rm el}$ describes the electron hopping (with amplitude $t$). 
 The second term describes the coupling to nonmagnetic and magnetic impurities, 
\begin{align}
 V_{\rm imp} &= 
	u_{\rm i} \sum_{i\in {\rm C}} c_{i}^\dagger c_i 
+   u_{\rm s} \sum_{j\in {\rm C'}} {\bm S}_j^{\rm imp} \cdot c_j^\dagger {\bm\sigma} \, c_j  , 
\end{align}
where $u_{\rm i}$ and $u_{\rm s}$ are the strengths of the coupling,
and ${\rm C}$ and ${\rm C'}$ are the sets of positions of nonmagnetic and magnetic impurities, 
respectively.
 The number of impurities on A and B sublattices are assumed equal.

 The coupling of conduction electrons to the localized spins is introduced by $H_{\rm sd}$,  
where ${\bm \sigma}  = {}^t (\sigma_x,\,\sigma_y,\,\sigma_z)$ are Pauli matrices 
and $J_{\rm sd}$ is the coupling constant. 
 With $\tilde {\bm l}$ and ${\bm n}$, its density is written as 
\begin{align}
  {\cal H}_{\rm sd}  
= - \frac{M}{s_n} \, ( \, \tilde {\bm l } \cdot \hat {\bm \sigma}_l + {\bm n} \cdot  \hat {\bm \sigma}_n ) , 
\label{eq:H_sd}
\end{align}
where $\hat {\bm \sigma}_l$ ($\hat {\bm \sigma}_n$) is the uniform (staggered) spin density of electrons, 
$s_n=2\hbar S/(2a^2)$, and $M = J_{\rm sd}S$.

We introduce a local unitary transformation in the electron spin space 
 that sorts the N\'eel vector at each site $i$ to the $\hat z$ direction
 \cite{Kohno2007,Shibata2011}, 
\begin{equation}
	U_i^\dagger (\bm n_i\cdot {\bm \sigma} ) \, U_i = \sigma^z
.
\end{equation}
We define an $SU(2)$ spin gauge field $A_{ij}$ by 
\begin{equation}
	U_{i}^\dagger U_{j} = e^{iA_{ij}}  . 
\end{equation}
 This is associated with the hopping from site $j$ to $i$, and satisfies $A_{ij} = -A_{ji}$. 
 It contains information about the texture of the N\'eel vector. 
 We assume slow spatial variation for ${\bm n}_i$, therefore, $A_{ij}$ is small 
and can be treated perturbatively. 
 We define the Fourier component $A_{\mu}({\bm q})$ by
\begin{align}
    A_{ij} &= \sum_{\bm q} A_{\mu}({\bm q}) \, e^{i\bm q\cdot(\bm r_i+\bm r_j)/2} , 
\end{align}
where $\hat \mu = \bm r_j - \bm r_i$,  
and expand it with Pauli matrices, 
\begin{align}
 A_{\mu}({\bm q}) 
&=  \sum_{\alpha = x,y,z} A_{\mu}^\alpha ({\bm q}) \, \frac{\sigma^\alpha}{2}  
  \equiv  {\bm A}_\mu \cdot  \frac{{\bm \sigma}}{2}  . 
\end{align}
 The corresponding $3\times 3$ matrix $\mathcal{R}_i$ is defined by 
\begin{align}
 U_{i}^\dagger \sigma^\alpha U_{i} = \mathcal{R}_i^{\alpha\beta} \sigma^\beta  .
\end{align}
 To first order in the spin gauge field,  the Hamiltonian in the Fourier representation 
is written as \cite{Nakane2020,nakazawa_SHE}
\begin{align}
 \tilde{H}_{\rm el}  
&=   \sum_{\bm k} \psi_{\bm k}^\dagger \varepsilon_{\bm k} \tau_1  \psi_{\bm k}
	+ \frac{1}{2} \sum_{i=x,\,y}\sum_{\bm k,\bm q} (\partial_i \varepsilon_{\bm k}) 
	  (\psi_{\bm k_+}^\dagger \sigma^\alpha \tau_1 \, \psi_{\bm k_-}) A_{i}^\alpha ({\bm q}) 
      + \tilde{H}_{\rm imp}  , 
\\
	\tilde{H}_{\rm sd} 
&= - M \sum_{\bm k} \psi_{\bm k}^\dagger \sigma^z \tau_3  \psi_{\bm k}  , 
\end{align}
 where $\tau_n$ and $\sigma^\alpha$ are Pauli matrices that act in sublattice (A or B) and spin 
($\uparrow$ or $\downarrow$) spaces, respectively, $\varepsilon_{\bm k} = -2t (\cos k_x + \cos k_y)$, 
$\bm k_{\pm} = \bm k\pm\frac{\bm q}{2}$,  and we set the lattice constant to unity. 
The electron creation operator in the sublattice representation is now given by
$\psi_{\bm k}^\dagger
= ( \tilde{c}^\dagger_{{\rm A}\bm k\uparrow}   ,
     \tilde{c}^\dagger_{{\rm A}\bm k\downarrow} ,
     \tilde{c}^\dagger_{{\rm B}\bm k\uparrow}   ,
     \tilde{c}^\dagger_{{\rm B}\bm k\downarrow} ) $, 
where
$\tilde{c}^\dagger_{{\rm A/B}\bm k\sigma}$ 
is the Fourier representation of the electron creation operator in the rotated frame 
with spin $\sigma$ and on sublattice ${\rm A/B}$.
 The $\bm k$-integral is taken in the reduced Brillouin zone, $|k_x+k_y| \leq \pi$. 
 Note that in $\tilde{H}_{\rm sd}$, the coupling to $\bm l$ is dropped 
since it only gives higher order terms in $\tilde {\bm l}$.

\section{Calculation of spin torques}

\subsection{Formalism}

We calculate the conduction electron spin densities 
$\tilde{\bm\sigma}\tau_\beta =  \sum_{\bm k}\psi^\dagger_{\bm k} {\bm\sigma}\tau_\beta \psi_{\bm k}$
in response to an applied electric field $E_i$
using the Kubo formula. 
We dropped the wave number dependence of the spin density operator
 since the spin gauge field already contains a spatial derivative.
The Kubo formula is given by
\begin{align}
&  \langle O \rangle_{\rm ne}
 = \lim_{\omega\to 0}
    \frac{ K_i (\omega+i0) - K_i (0) }{i\omega} E_i  , 
\\
&  K_i (i\omega_\lambda)
 =  \int_0^\beta d\tau e^{i\omega_\lambda \tau}
    \langle T_\tau  O (\tau)   {\tilde J}_i \rangle  , 
\end{align}
where $O = \sigma^\alpha$ or $\sigma^\alpha \tau_3$. 
 The current operator in the rotated frame is given to first order in spin gauge field by
\begin{align}
	\tilde{J}_i
&= -e\sum_{\bm k}
	 \psi_{\bm k}^\dagger 
	 [ (\partial_i \varepsilon_{\bm k}) \tau_1 ]
	 \psi_{\bm k} 
  -e \sum_{\bm k,\bm q} 
	\psi_{\bm k+ \frac{\bm q}{2}}^\dagger 
	[  (\partial_i ^2\varepsilon_{\bm k}) \sigma^\alpha  \tau_1 ]
	\psi_{\bm k- \frac{\bm q}{2} }  A_{i}^\alpha ({\bm q}) . 
\end{align}
 The average $\langle\cdots\rangle$ is taken in the thermal equilibrium state, determined by
$\tilde{H}_{\rm el}+\tilde{H}_{\rm sd}$,
to first order in the spin gauge field.

\begin{figure}[tbp]
    \centering
    \includegraphics[width=1.\textwidth]{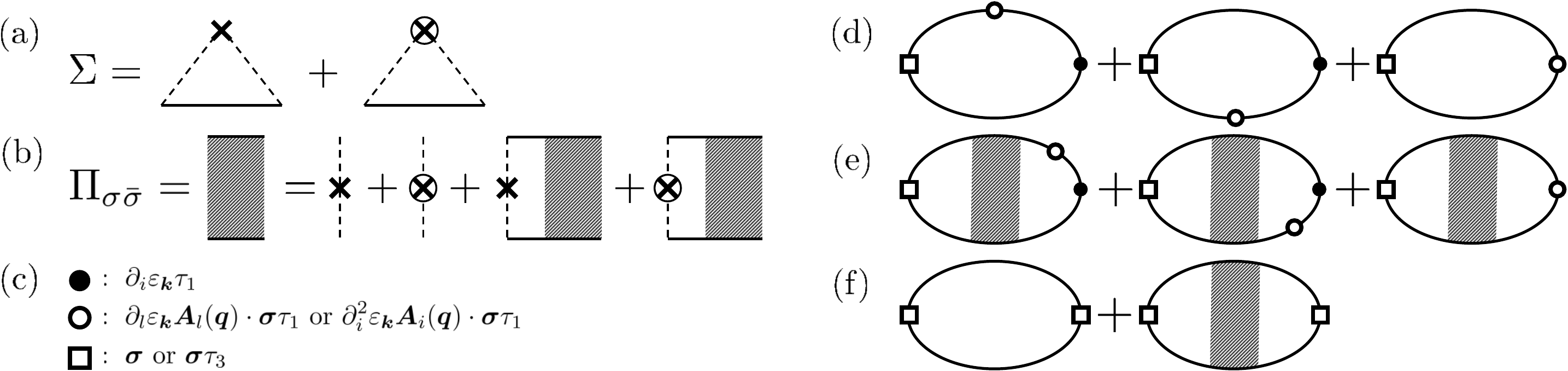}
    \caption{
     Feynman diagrams for the current-induced spin torques and Gilbert damping. 
     (a) and (b) show the treatment of random impurities, and (c)-(f) show the response functions 
(electron spin density in response to the electric field ${\bm E}$ 
or time-dependent uniform/staggered magnetization, ${\bm n}$ or $\tilde {\bm l}$). 
     The solid line represents the electron Green's function, 
     and the dashed line with a cross (circled cross) represents nonmagnetic (magnetic) impurity scattering.  
        (a) Self-energy in the Born approximation. 
        (b) Four-point vertex with impurity ladder. 
        The upper (lower) electron line is in the retarded (advanced) branch and has spin $\sigma$ ($\bar\sigma$).
        (c) Definition of vertices. 
    The filled circle represents the current vertex $(\partial_i \varepsilon_{\bm k}) \tau_1$, where the direction $i$ is coupled to the external electric field $E_i$. 
    The empty circles represent vertices that contain the spin gauge field ${\bm A}$ 
     coming either from the perturbation Hamiltonian 
    $(\partial_l \varepsilon_{\bm k}) \bm A_l ({\bm q})\cdot {\bm \sigma} \tau_1$ 
    or from the current vertex 
    $(\partial^2_i \varepsilon_{\bm k}) {\bm A}_i ({\bm q}) \cdot {\bm \sigma} \tau_1$. 
    The empty square is the uniform ($\bm\sigma$) or staggered (${\bm\sigma} \tau_3$) spin density. 
    Diagrams in (d) and (e) represent current-induced torques, in which 
    the right vertex represents the charge current that couples to ${\bm E}$. 
        (d) Diagrams first order in ${\bm A}$.
        The so-called Fermi-sea terms, consisting of only retarded or advanced Green's functions, 
       also need to be retained for the staggered spin density. 
        (e) Diagrams first order in ${\bm A}$ with ladder vertex corrections. 
    Diagrams in (f) represent Gilbert damping. 
    }
\label{fig:fig1}
\end{figure}

The calculation is executed using the Green's function
 of the tight-binding electrons in a homogeneous antiferromagnetic state.
The retarded Green's function is given by
\begin{align}
 G^R_{\bm k} = \mu^R_{\bm k} + T^R_{\bm k}\tau_1 + J^R_{\bm k} \sigma^z \tau_3 ,
\end{align}
where $\mu^R = (\mu+i\gamma_0)/D^R_{\bm k}$,
 $T^R = \varepsilon_{\bm k}/D^R_{\bm k}$, 
 $J^R  = (-M +i\gamma_3)/D^R_{\bm k}$, and
$D^R_{\bm k} = (\mu+i\gamma_0)^2 - \varepsilon_{\bm k}^2-(M-i\gamma_3)^2$.
The damping constants are calculated in the Born approximation 
(see Fig.~\ref{fig:fig1} (a)) as 
$\gamma_0  = \gamma_{\rm n} + \gamma_z + 2\gamma_\perp$ and
$\gamma_3  = (\gamma_{\rm n} + \gamma_z - 2\gamma_\perp) M/\mu$, 
with $\gamma_{\rm n}=\pi n_{\rm i}u_{\rm i}^2 \nu$ 
 from nonmagnetic impurities and
\begin{align}
  \gamma_\perp = \pi n_{\rm s} u_{\rm s}^2 \, \overline{S_\perp^2} \, \nu  , 
\ \ \ 
  \gamma_z &= \pi n_{\rm s} u_{\rm s}^2 \, \overline{S_z^2} \, \nu , 
\end{align}
from magnetic impurities.
 We wrote $n_{\rm i}$ ($n_{\rm s}$) for the nonmagnetic (magnetic) impurity concentration,
and we took the directional average of magnetic impurity spins in the rotated frame,  
$\tilde{\bm S}^{\rm imp} = \mathcal{R}^{-1}{\bm S}^{\rm imp}$, as
$\langle \tilde{S}^{\rm imp}_\alpha \rangle_{\rm imp} = 0$,
$\langle (\tilde{S}^{\rm imp}_{j, z})^2 \rangle_{\rm imp} \equiv \overline{S_z^2}$,
and 
$\langle ( \tilde{S}^{\rm imp}_{j, x})^2 \rangle_{\rm imp} 
= \langle ( \tilde{S}^{\rm imp}_{j, y})^2 \rangle_{\rm imp} 
\equiv \overline{S_\perp^2}$. 
The density of states is given by
 $\nu = \nu (\mu)  = \frac{1}{N} \sum_{\bm k} \delta(|\mu|-E_{\bm k})$ 
where
$ \pm E_{\bm k} = \pm \sqrt{\varepsilon_{\bm k}^2+M^2} $, 
and $\mu$ is the chemical potential.

 To work in consistency with the Born approximation, 
 we consider the impurity-ladder vertex correction as shown in Fig.~\ref{fig:fig1} (b). 
 The relevant correction comes from retarded and advanced Green's functions with opposite spins, 
and is obtained as 
\begin{comment}
\begin{align}
    \Pi_{\sigma\bar{\sigma}}
&= \frac{2}{\pi\nu} \frac{4\gamma^2\mu^2}{\mu^2-M^2}
    \bigg[ Dq^2 - i\omega
\nonumber\\
&\qquad
   + 2 \frac{\gamma_{\rm n} M^2+\gamma_z\mu^2+\gamma_\perp(\mu^2-M^2)
        }{ (\gamma_{\rm n}-\gamma_z)(\mu^2-M^2) \tau }
    \bigg]^{-1}
\\
&\simeq
    \frac{2\gamma}{\pi\nu}
    \cdot
    \frac{\gamma_{\rm n}-\gamma_z}{
        \frac{M^2}{\mu^2}\gamma_{\rm n}
        + \gamma_z
        +(1-\frac{M^2}{\mu^2})\gamma_\perp }
\end{align}
\end{comment}
\begin{align}
    \Pi_{\sigma\bar{\sigma}}
&= \frac{2}{\pi\nu} \frac{4\gamma^2\mu^2}{\mu^2-M^2} \, 
    \frac{1}{Dq^2 - i\omega + \tau_\varphi^{-1} + \tau_{\rm s}^{-1} }  
\label{eq:Pi_0}
\\
&\simeq
    \frac{2\gamma}{\pi\nu}
    \cdot
    \frac{\gamma_{\rm n} - \gamma_z}{
        \frac{M^2}{\mu^2} \gamma_{\rm n}
        + \gamma_z
        +(1-\frac{M^2}{\mu^2}) \gamma_\perp }   , 
\label{eq:Pi_1}
\end{align}
where $\gamma = \gamma_0 + (M/\mu) \gamma_3$ is the electron scattering rate,  
\begin{align}
  \frac{1}{\tau_\varphi}  
&= \frac{4M^2}{\mu^2 } \left[ 
     \frac{\mu^2+M^2}{\mu^2-M^2} \, \gamma_{\rm n} 
    + 3\gamma_\perp + \frac{2(2\mu^2+M^2)}{\mu^2-M^2} \, \gamma_z  \right]  ,
\\
  \frac{1}{\tau_{\rm s}}  &= 4 \, (\gamma_\perp + \gamma_z  ) , 
\end{align}
are the spin-dephasing and spin-relaxation rates, respectively. 
 We decompose the former into 
\begin{align}
  \frac{1}{\tau_{\varphi 0}}  
&= \frac{4M^2}{\mu^2 }  
     \frac{\mu^2+M^2}{\mu^2-M^2} \, \gamma_{\rm n}   ,
\\
  \frac{1}{\tau_{\varphi 1}}  
&= \frac{4M^2}{\mu^2 } \left[ 3\gamma_\perp + \frac{2(2\mu^2+M^2)}{\mu^2-M^2} \, \gamma_z  \right]  . 
\end{align}

%With the Green's function and vertex correction in hand, we can now calculate the current-induced spin densities, $\langle {\bm \sigma}_l \rangle$ and $\langle {\bm \sigma}_n \rangle$ 
%(notation in the previous section), or equivalently,  
%$\langle {\bm \sigma} \rangle$ and $\langle {\bm \sigma} \tau_3 \rangle$ 
%(notation in the present section), in textured antiferromagnets. 
% We calculate the Feynman diagrams shown in Figs.~\ref{fig:fig1} (d) and (e). 
%%, which are processes without and with vertex corrections, respectively. 
%% and the latter are terms with vertex correction.
%The spin densities are calculated to leading order in spatial gradient (i.e., first order in the spin gauge field), 
%and the electron damping $\gamma$. 
% We assume weak spin relaxation, $\gamma_z,  \gamma_\perp \ll \gamma_{\rm n}$, 
%and consider its lowest nontrivial order. 
%% Details are described in Appendix. 

\subsection{Uniform spin density}

 The uniform spin density is finite without magnetic impurities,
so magnetic impurities are not considered here.
 The first two diagrams in Fig.~\ref{fig:fig1} (d) give
\begin{align}
    \text{(d1) + (d2)}
    &=
    \frac{1}{i\omega} \frac{-\omega}{2\pi i}  \frac{1}{2}
    {A}_i^\alpha  \sum_{\bm k} (\partial_i\varepsilon_{\bm k})^2
    {\rm tr} ( {\bm \sigma}^\perp G^R \sigma^\alpha\tau_1 G^R \tau_1 G^A ) + {\rm c.c.}
\nonumber \\
&=
    {\bm A}_i^\perp
    \frac{2 \nu \tau}{\mu}
    \bigg[
        \frac{-2M^2}{\mu^2+M^2}
         \langle\!\langle (\partial_i\varepsilon_{\bm k})^2\rangle\!\rangle
        -\langle\!\langle
        \varepsilon_{\bm k}(\partial_i^2\varepsilon_{\bm k})
        \rangle\!\rangle
    \bigg]  , 
\end{align}
where ${\bm A}_i^\perp = {\bm A}_i - A_i^z \hat z$ and 
$\langle\!\langle \cdots \rangle\!\rangle 
= \nu^{-1} \frac{1}{N} \sum_{\bm k}(\cdots)\delta(|\mu|-E_{\bm k})$.
 The third diagram (d3) gives
\begin{align}
    \text{(d3)}
&=  \frac{1}{i\omega}  \frac{-\omega}{2\pi i}  \frac{1}{2}  A_i^\alpha
    \sum_{\bm k}  (\partial_i^2\varepsilon_{\bm k}) \, 
    {\rm tr} \left[ {\bm \sigma}^\perp G^R \sigma^\alpha\tau_1 G^A  \right] 
\nonumber \\
&= {\bm A}_i^\perp  \frac{2 \nu \tau}{\mu}
    \langle\!\langle   \varepsilon_{\bm k}(\partial_i^2\varepsilon_{\bm k}) \rangle\!\rangle  , 
\end{align}
which cancels the term 
$ \sim \langle\!\langle
        \varepsilon_{\bm k}(\partial_i^2\varepsilon_{\bm k})
    \rangle\!\rangle $
in (d1)$+$(d2). 
 In Fig.~\ref{fig:fig1} (e), the first two diagrams, (e1) and (e2), give
\begin{align}
    \text{(e1) + (e2)}
    &=
    \frac{1}{4^2}
    \frac{1}{i\omega}
    \frac{-\omega}{2\pi i}
    \frac{1}{2}
    {A}_i^\alpha
    \sum_{\bm k'}
    {\rm tr} \left[  \sigma^\perp G^R  \sigma^\perp G^A \right] 
    \Pi_{\sigma\bar{\sigma}}
    \biggl\{ 
    \sum_{\bm k} (\partial_i\varepsilon_{\bm k})^2
    {\rm tr}  (  {\sigma^\perp} G^R \sigma^\alpha \tau_1 G^R  \tau_1 G^A )
    + {\rm c.c.}
    \biggr\} 
\nonumber \\
&=  {\bm A}_i^\perp
    \frac{2 \nu \tau_{\varphi 0} }{\mu}
    \bigg[
        \frac{-2M^2}{\mu^2+M^2}
         \langle\!\langle (\partial_i\varepsilon_{\bm k})^2\rangle\!\rangle
        - \langle\!\langle
        \varepsilon_{\bm k}(\partial_i^2\varepsilon_{\bm k})
        \rangle\!\rangle
    \bigg]  , 
\end{align}
where $\Pi_{\sigma\bar{\sigma}}$ is the ladder vertex part [Eq.~(\ref{eq:Pi_1})], 
and we defined $\frac{\mu^2-M^2}{2M^2} \tau = \tau_{\varphi 0}$.
 The third diagram (e3) can be similarly calculated, 
\begin{align}
    \text{(e3)} 
    &=
    {\bm A}_i^\perp
    \frac{2 \nu \tau_{\varphi 0} }{\mu}
    \langle\!\langle
        \varepsilon_{\bm k}(\partial_i^2\varepsilon_{\bm k})
    \rangle\!\rangle
.
\end{align}

Thus, the total uniform spin density in the rotated frame is obtained as 
\begin{align}
    \langle \tilde{\bm \sigma}^\perp \rangle_{\rm ne} =
    \text{Fig.~\ref{fig:fig1} (d)+(e)}
    &=
    {\bm A}_i^\perp
    (\tau + \tau_{\varphi 0})
    \frac{2 \nu}{\mu}
    \frac{-2M^2}{\mu^2+M^2}
     \langle\!\langle (\partial_i\varepsilon_{\bm k})^2\rangle\!\rangle (-eE_i) 
\nonumber \\ 
&= {\bm A}_i^\perp  \tau  \frac{2 \nu}{\mu}
     \langle\!\langle (\partial_i\varepsilon_{\bm k})^2\rangle\!\rangle  e E_i 
\nonumber \\ 
&=  {\bm A}_i^\perp  \frac{2\mu}{ \mu^2-M^2 } \,  \frac{j_i}{2e}  , 
\end{align}
where 
${\bm j} = 2e^2 D\nu {\bm E} 
 = 2e^2  \frac{\mu^2-M^2}{\mu^2} 
 \langle\!\langle (\partial_i\varepsilon_{\bm k})^2 \rangle\!\rangle \, \nu \tau {\bm E} $ 
is the electric current density \cite{Nakane2020}. 
 The spin density in the laboratory frame is obtained from 
${\cal R} {\bm A}_i^\perp = - {\bm n} \times \partial_i {\bm n}$, 
\begin{align}
   \langle {\bm \sigma}^\perp \rangle_{\rm ne}  
= {\cal R}  \langle \tilde{\bm \sigma}^\perp \rangle_{\rm ne} 
&= - {\bm n} \times \partial_i {\bm n} \frac{2 \mu}{ \mu^2-M^2 } \,  \frac{j_i}{2e}  . 
\end{align}

\subsection{Staggered spin density}

 The staggered spin density vanishes without magnetic impurities, so they must be considered.
 Note that since the staggered spin density is zeroth order in the scattering time $\tau$, 
the so-called ``Fermi-sea terms" that contain only advanced or retarded Green's functions 
need to be retained. 

 Let us first calculate the Fermi-surface terms. 
 The first two diagrams in Fig.~\ref{fig:fig1} (d) without vertex correction are calculated as 
\begin{align}
    \text{ [(d1) + (d2)}]^{\text{surf}}
    &=
    \frac{1}{i\omega}
    \frac{-\omega}{2\pi i}
    \frac{1}{2}
    {A}_i^\alpha
    \sum_{\bm k}  (\partial_i\varepsilon_{\bm k})^2  \, 
    {\rm tr} \left[ 
    {\sigma^\perp \tau_3} G^R \sigma^\alpha\tau_1 G^R \tau_1 G^A \right] 
    + {\rm c.c.}
\nonumber \\ 
&=   {\bm A}_i \times{\hat z}
    \sum_{\bm k} 2\delta(\mu^2-E_{\bm k}^2)
    \frac{\tau }{|\mu|}
    \biggl\{ 
    %%%%%%%%%%%%%%%%%%%%%%%%%%%%%
    - (\partial_i\varepsilon_{\bm k})^2\gamma_3
    + (\mu\gamma_3 + M \gamma_0)
    \biggr[  2(\partial_i\varepsilon_{\bm k})^2  \gamma_0\frac{\tau}{\mu}
   - \mu
   \frac{\varepsilon_{\bm k}(\partial_i^2\varepsilon_{\bm k})-(\partial_i\varepsilon_{\bm k})^2}{\mu^2-M^2}
    \biggr] 
    %%%%%%%%%%%%%%%%%%%%%%%%%%%%%
    \biggr\}  . 
\end{align}
 The anomalous velocity term (the third diagram) without vertex correction is calculated as 
\begin{align}
    \text{(d3)}^{\text{surf}}
    &=
    \frac{1}{i\omega}   \frac{-\omega}{2\pi i}  \frac{1}{2}  A_i^\alpha
    \sum_{\bm k} (\partial_i^2\varepsilon_{\bm k})
    {\rm tr}  (  \sigma^\perp\tau_3 G^R  \sigma^\alpha\tau_1 G^A )
\nonumber \\ 
&=  {\bm A}_i\times{\hat z}
    \sum_{\bm k} (\partial_i^2\varepsilon_{\bm k})  \varepsilon_{\bm k}
    2 \gamma_3 \frac{\tau}{|\mu|}\delta(\mu^2-E_{\bm k}^2)  . 
\end{align}
The Fermi-sea terms in (d1)$+$(d2) are given by
\begin{align}
    \text{ [(d1) + (d2)}]^{\text{sea}}
    &=
    \frac{1}{i\omega}  \frac{1}{2\pi i} \frac{1}{2}  {A}_i^\alpha
    \sum_{\bm k}  (\partial_i\varepsilon_{\bm k})^2
    \int d\varepsilon
    \biggl\{ 
       f (\varepsilon_-) \, {\rm tr} \left[\, \sigma^\perp\tau_3 G^R_+  \, 
            ( \tau_1 G^R_-  \sigma^\alpha\tau_1 +  \sigma^\alpha\tau_1 G^R_+  \tau_1)  \,  G^R_-  \right] 
\nonumber\\
&\hskip 45mm
    -  f (\varepsilon_+) \, {\rm tr} \left[\, \sigma^\perp\tau_3 G^A_+   \, 
            ( \tau_1  G^A_-  \sigma^\alpha\tau_1 + \sigma^\alpha\tau_1  G^A_+  \tau_1) \,  G^A_-  \right] 
    \biggr\}    
\nonumber \\ 
&=
    {\bm A}_i\times{\hat z}  \, 
    \frac{ M \, ({\rm sgn} \mu )  }{\mu^2-M^2}
    \sum_{\bm k}
    \bigl\{ \varepsilon_{\bm k}(\partial_i^2\varepsilon_{\bm k})
     - (\partial_i\varepsilon_{\bm k})^2 \bigr\} 
    \delta(\mu^2-E_{\bm k}^2)  , 
\end{align}
where 
$f (\varepsilon) = \theta (-\varepsilon)$ is the Fermi-Dirac distribution function 
at zero  temperature, $\varepsilon_\pm = \varepsilon \pm \omega/2$, 
and $G^{R/A}_\pm = G^{R/A}(\varepsilon\pm\omega/2)$.

The Fermi-sea term from the anomalous velocity diagram vanishes,  
\begin{align}
    \text{(d3)}^{\text{sea}}
    &=
    \frac{1}{2\pi i}
    \frac{1}{2}
    {A}_i^\alpha
    \sum_{\bm k}
    (\partial_i^2\varepsilon_{\bm k})
    \int d\varepsilon
    \bigg\{
    f (\varepsilon_-) \, 
    {\rm tr} \left[ \, \sigma^\perp \tau_3  G^R_+ \sigma^\alpha\tau_1 G^R_-  \right] 
    -
    f (\varepsilon_+) \, 
    {\rm tr} \left[ \, \sigma^\perp \tau_3  G^A_+ \sigma^\alpha\tau_1 G^A_-  \right] 
    \bigg\}    
\nonumber \\ 
&=0  .
\end{align}
 Adding up the Fermi-surface and Fermi-sea terms, we obtain 
\begin{align}
    \text{ [(d1) + (d2) + (d3)}]^{\text{surf \& sea}}
    &=
    {\bm A}_i \times{\hat z} \, 
    \frac{2 \nu}{\mu}
    \langle\!\langle (\partial_i\varepsilon_{\bm k})^2\rangle\!\rangle
    \frac{2\tau^2\gamma_0}{\mu^2}
    (\mu\gamma_3 + M\gamma_0)  . 
\end{align}
 Next, the diagrams with vertex corrections, shown in Fig.~\ref{fig:fig1} (e), give
\begin{align}
    \text{(e1) + (e2) + (e3)}
    &=
    \frac{1}{i\omega}
    \frac{-\omega}{2\pi i}
    \frac{1}{4^2}
    \frac{1}{2}
    A_i^\alpha
    \bigg(
    \sum_{\bm k}
    {\rm tr}
    (\sigma^\perp\tau_3
    G^R
    \sigma^\lambda
    G^A
    )
    \bigg)
    \frac{2\gamma}{\pi\nu}
    \cdot
    \frac{\gamma_{\rm n}-\gamma_z}{
        \frac{M^2}{\mu^2}\gamma_{\rm n}
        + \gamma_z
        + (1-\frac{M^2}{\mu^2})\gamma_\perp
    }
\nonumber\\
&\quad  \times 
    \biggl\{  \sum_{\bm k}
    (\partial_i\varepsilon_{\bm k})^2 \, 
    {\rm tr} \left[ \sigma^\lambda G^R  \sigma^\alpha  G^R \tau_1 G^A  \right] 
    + {\rm c.c.}
    + (\partial_i^2\varepsilon_{\bm k}) \, 
    {\rm tr}  \left[ \sigma^\lambda G^R \sigma^\alpha\tau_1 G^A  \right] 
    \biggr\}   . 
\end{align}
 The first trace is evaluated as 
\begin{align}
    \sum_{\bm k}
    {\rm tr} \left[  \sigma^\perp\tau_3 G^R \sigma^\lambda G^A  \right] 
    &=
    -i \,
    {\rm tr}
    (\sigma^\perp\sigma^\lambda\sigma^z )
    (\gamma_3\mu+\gamma_0 M)
    \frac{2\pi\tau\nu}{\mu^2}
.
\end{align}
The second trace is the same as calculation of the uniform spin density. 
 Thus, we have 
\begin{align}
    \text{(e1) + (e2) + (e3)}
&=  {\bm A_i}\times{\hat z} \, 
    \frac{2 \nu}{\mu}
    \langle\!\langle (\partial_i\varepsilon_{\bm k})^2\rangle\!\rangle
    \frac{\gamma_{\rm n}-\gamma_z}{
        \frac{M^2}{\mu^2}\gamma_{\rm n}
        +\gamma_z
        +(1-\frac{M^2}{\mu^2})\gamma_\perp
    }
    \bigg(
        \frac{-M (\mu\gamma_3 + M \gamma_0)^2}{4\mu^4\gamma^2}
    \bigg)  . 
\end{align}
 Therefore, the total staggered spin density in the rotated frame is obtained as 
\begin{align}
  \langle \tilde{\bm \sigma}^\perp \tau_3 \rangle_{\rm ne} 
&= \text{ [(d1) + (d2) + (d3)}]^{\text{surf \& sea}} + \text{(e1) + (e2) + (e3)}
\nonumber \\ 
&=
    {\bm A_i}\times{\hat z} \, 
    \frac{2 \nu \tau}{\mu}
    \langle\!\langle (\partial_i\varepsilon_{\bm k})^2\rangle\!\rangle
    (\gamma_\perp+\gamma_z)
    \frac{
        2M (\gamma_{\rm n} + \gamma_z)
    }{
        M^2\gamma_{\rm n}
        +\mu^2\gamma_z
        +(\mu^2-M^2) \gamma_\perp
    }  (-eE_i) 
\nonumber \\ 
&=
    {\bm A_i}\times{\hat z} \, 
    \frac{2 \nu \tau}{\mu}
    \langle\!\langle (\partial_i\varepsilon_{\bm k})^2\rangle\!\rangle
    \frac{ 2 (\gamma_\perp+\gamma_z) }{M}  (-eE_i) 
\nonumber \\ 
&= -  {\bm A_i}\times{\hat z} \, 
    \frac{ 2 (\gamma_\perp+\gamma_z) }{M}  
   \frac{2 \mu}{ \mu^2-M^2 } \,  \frac{j_i}{2e}  , 
\end{align}
where in the third equality, 
we retained lowest-order terms in spin relaxation (magnetic impurities). 
 In the laboratory frame, we have 
\begin{align}
   \langle {\bm \sigma}^\perp \tau_3 \rangle_{\rm ne} 
= {\cal R} \langle \tilde{\bm \sigma}^\perp \tau_3 \rangle_{\rm ne} 
&=  \beta_n \,  \partial_i {\bm n} \,  
   \frac{2 \mu}{ \mu^2-M^2 } \,  \frac{j_i}{2e}  , 
\end{align}
where $\beta_n = 2 (\gamma_\perp + \gamma_z) /M$. 

\subsection{Gilbert damping for staggered moment}

To calculate the damping term due to conduction electrons,
we use the small amplitude method and consider 
\begin{align}
    \mathcal{H}^{\delta n}_{\rm sd} 
&=
    -M
    \sum_{\bm k} 
    \psi^\dagger_{\bm k} {\bm\sigma}\tau_3\psi_{\bm k}\cdot\delta{\bm n}
\end{align}
to be the perturbing Hamiltonian, 
with the dynamic deviation of the N\'eel vector $\delta{\bm n}$ in the $xy$ plane \cite{Kohno2006}.
 We calculate the $\omega$-linear terms of the staggered spin density 
 in response to $\mathcal{H}^{\delta n}_{\rm sd}$
using the Kubo formula
\begin{align}
    \langle {\bm \sigma} \tau_3 (\omega) \rangle_{\rm ne}^{\delta n}
    &=
    -\frac{i}{\hbar}
    \int_{-\infty}^\infty
    e^{i\omega t}
    \theta(t)
    \langle
    [
    \hat{\bm \sigma}_n (t), 
    \mathcal{H}^{\delta n}_{\rm sd}(0)
    ]
    \rangle
    \, dt
.
\end{align}

The terms without vertex correction is given by
\begin{align}
    \langle {\bm\sigma} \tau_3 \rangle^{\delta n}
    &=
    -M\delta{n}^\alpha
    \frac{1}{2\pi i}
    \int d\varepsilon
    \bigg[
    \left(
    -f(\varepsilon_+)
    +f(\varepsilon_-)
    \right)
    {\bm \sigma}\tau_3 G^R(\varepsilon_+) 
     {\sigma^\alpha\tau_3}G^A(\varepsilon_-)
    \nonumber\\&\qquad\qquad
    +f(\varepsilon_+)
    {\bm \sigma}\tau_3 G^A(\varepsilon_+) 
     {\sigma^\alpha\tau_3}G^A(\varepsilon_-)
    -f(\varepsilon_-)
    {\bm \sigma}\tau_3 G^R(\varepsilon_+) 
     {\sigma^\alpha\tau_3}G^R(\varepsilon_-)
    \bigg] .
\end{align}
The RA term gives
\begin{align}
    \langle {\bm\sigma} \tau_3 \rangle_{RA}^{\delta n}
    &=
    -M\delta{\bm n}
    \frac{\omega}{2\pi i}
    4(\mu^R\mu^A -T^RT^A - J^RJ^A)
    \\&=
    2i\omega
    M\delta{\bm n}
    \frac{2 \nu \tau}{\mu^2}
    (\gamma_0^2-\gamma_3^2)
\end{align}
The RR and AA terms give
\begin{align}
    \langle {\bm\sigma} \tau_3 \rangle_{RR}^{\delta n}
    +\langle {\bm\sigma} \tau_3 \rangle_{AA}^{\delta n}
    &=
    \frac{1}{2}
    M\delta{n}^\alpha
    \frac{\omega}{2\pi i}
    \bigg(
    {\bm \sigma}\tau_3 G^R
     {\sigma^\alpha\tau_3}G^R
    +
    {\bm \sigma}\tau_3 G^A
     {\sigma^\alpha\tau_3}G^A
    \bigg)
    \\&=0
\end{align}
With vertex correction the RA term is given by,
\begin{align}
    \langle {\bm\sigma} \tau_3 \rangle_{RA,V}^{\delta n}
    &=
    -
    \frac{1}{4^2}
    M\delta{n}^\alpha
    \frac{\omega}{2\pi i}
    {\rm tr}(
    {\bm \sigma}\tau_3 G^R
    \sigma^\beta
    G^A
    )
    \,
    \Pi
    \,
    {\rm tr}(
    \sigma^\beta
    G^R
     {\sigma^\alpha\tau_3}
    G^A
    )
    \\&=
    -i\omega
    M\delta{\bm n}
    (
    \gamma_3\mu
    +\gamma_0M
    )^2
    \frac{2 \nu \tau}{\mu^2}
    \frac{\gamma_{\rm n}-\gamma_z}{M^2\gamma_{\rm n} + \mu^2\gamma_z + (\mu^2-M^2)\gamma_\perp}
\end{align}
Lastly, let us consider the vertex correction on the RR and AA terms. 
\begin{align}
    \langle {\bm\sigma} \tau_3 \rangle_{RR,V}^{\delta n}
    +\langle {\bm\sigma} \tau_3 \rangle_{AA,V}^{\delta n}
    &=
    \frac{1}{2\cdot 4^2}
    M\delta{n}^\alpha
    \frac{\omega}{2\pi i}
    \bigg(
    {\rm tr}[
    {\bm \sigma}\tau_3 G^A
     \sigma^\beta\tau_3G^A
    ]
    \,\Pi_1\,
    {\rm tr}[
    \sigma^\beta\tau_3
    G^A
     {\sigma^\alpha\tau_3}G^A
    ]
    +
    {\rm c.c.}
    \bigg)
    \\&=
    M\delta{\bm n}
    i\omega
    (\gamma_{\rm n}-\gamma_z)
    \frac{2\nu}{\mu^2} 
\end{align}
where $\Pi_1$ is the vertex correction with one non-magnetic and magnetic impurity, 
$\Pi_1 = \frac{4}{\pi\nu}(\gamma_{\rm n}-\gamma_z)$.
Finally, adding up the terms obtained 
\begin{align}
    \langle {\bm\sigma} \tau_3 \rangle_{\rm ne}^{\delta n}
    &=
    \langle {\bm\sigma} \tau_3 \rangle_{RA}^{\delta n}
    +\langle {\bm\sigma} \tau_3 \rangle_{RR}^{\delta n}
    +\langle {\bm\sigma} \tau_3 \rangle_{AA}^{\delta n}
    +\langle {\bm\sigma} \tau_3 \rangle_{RA,V}^{\delta n}
    +\langle {\bm\sigma} \tau_3 \rangle_{RR,V}^{\delta n}
    +\langle {\bm\sigma} \tau_3 \rangle_{AA,V}^{\delta n}
    \\&=
    -2 \, \delta\dot{\bm n}
    \frac{2\nu}{M}
    \left[
        \gamma_z + \gamma_\perp
        + \frac{M^2}{\mu^2} 
        (
         \gamma_\perp - \gamma_z  
        )
    \right]
\end{align}
to the leading order in magnetic impurities.
%The Gilbert damping parameter is then given by
%\begin{align}
%    \alpha_n &=
%    \frac{\nu}{2s_n}
%    \left[
%        \gamma_z + \gamma_\perp
%        + \frac{M^2}{\mu^2} 
%        (
%        \gamma_\perp - \gamma_z  
%        )
%    \right]
%.
%\end{align}
This gives the damping parameter 
\begin{align}
    \alpha_n &= \left[
        \gamma_z + \gamma_\perp  + \frac{M^2}{\mu^2}  (  \gamma_\perp - \gamma_z )
    \right] \frac{2\nu}{s_n}
\end{align}
where $s_n = 2 \hbar S/(2a^2)$.

\subsection{Gilbert damping for uniform moment}

The uniform Gilbert damping parameter $\alpha_l$ is similarly calculated 
using the small amplitude method as
the $\omega$-linear terms of the uniform spin density
in response to the s-d coupling to $\tilde{\bm l}$. 
The result is given by
%\begin{align}
%    \langle {\bm \sigma}_l \rangle_{\rm na}
%    &=
%    -M\tilde {\bm l}
%    \frac{\omega}{2\pi i}
%    4(\mu^R\mu^A +T^RT^A - J^RJ^A)
%    \\&=
%    M\tilde {\bm l}
%    i\omega
%    2(\mu^2  - M^2)\frac{\tau\nu}{\mu^2}
%\end{align}
%\begin{align}
%    \langle {\bm \sigma}_l \rangle_{\rm na, V}
%    &=
%    -M{ l^\beta}
%    \frac{\omega}{2\pi i}
%    \frac{1}{4^2}
%    {\rm tr}[{\bm \sigma} G^R \lambda G^A ]
%    \Pi
%    {\rm tr}[\lambda G^R {\sigma^\beta} G^A]
%    \\&=
%    M\tilde {\bm l}
%    2i\omega
%    (\mu^2-M^2)\frac{\tau_\phi^0\nu}{\mu^2}
%\end{align}
\begin{align}
    \langle {\bm \sigma}  \rangle_{\rm na}
    &=
    2M
    i\omega\tilde {\bm l}
    (\mu^2  - M^2)\frac{2 (\tau+\tau_\phi^0)\nu}{\mu^2}
    \\&=
    i\omega\tilde {\bm l}
    \frac{(\mu^2+M^2)(\mu^2  - M^2)}{M}\frac{2 \nu\tau}{\mu^2}  , 
\end{align}
so 
\begin{align}
    \alpha_l
    &= \frac{(\mu^2+M^2)(\mu^2  - M^2)}{\mu^2} \frac{\nu\tau}{s_n} .
\end{align}